\documentclass[hidelinks,12pt]{article}

\usepackage[margin=1in]{geometry} 
\usepackage{graphicx}              
\usepackage{amsmath}               
\usepackage{mathtools} 
\usepackage{amsfonts}              
\usepackage{amsthm}                
\usepackage{threeparttable}
\usepackage{setspace}  
\usepackage{indent first}
\usepackage[font={footnotesize,singlespacing},labelfont=bf]{caption}
\usepackage[font={doublespacing},labelfont=bf]{caption}
\usepackage{lineno}
\usepackage{titlesec}
\usepackage{gensymb}
\usepackage{titling}
\usepackage{titlesec}
\usepackage{listings}
\usepackage{booktabs}
\usepackage{multirow}
\usepackage{lscape}
\usepackage{authblk}
\usepackage{appendix}
\usepackage{geometry}
\usepackage{hyperref}
\usepackage[sort&compress]{natbib}
\bibpunct{(}{)}{,}{a}{}{;}

{\large }

\newcommand{\eqname}[1]{\tag*{#1}}

\newcommand{\mtnote}[1]{\textsuperscript{\TPTtagStyle{#1}}} 

\lstdefinelanguage{Renhanced}[]{R}{%
   morekeywords={acf,ar,arima,arima.sim,colMeans,colSums,is.na,is.null,%
     mapply,ms,na.rm,nlmin,replicate,row.names,rowMeans,rowSums,seasonal,%
     sys.time,system.time,ts.plot,which.max,which.min},
   deletekeywords={c},
   alsoletter={.\%},%
   alsoother={:_\$}}
\titlespacing*{\section}
{0pt}{0ex plus 2ex minus .1ex}{0ex plus .1ex}
\titlespacing*{\subsection}
{0pt}{0ex plus .1ex minus .1ex}{0ex plus .1ex}
\titlespacing*{\subsubsection}
{0pt}{1ex plus 1ex minus .1ex}{1ex plus .1ex}

\newcommand{\indexfonction}[1]{\index{#1@\texttt{#1}}}

\titleformat*{\subsection}{ \bfseries}
\titleformat*{\subsubsection}{\itshape}



\usepackage{fancyhdr}
\begin{document}
\title{Competition among eggs shifts to cooperation along a sperm supply gradient in an external fertilizer}
\author[ ]{Daniel K. Okamoto}
\affil[ ]{Department of Ecology, Evolution and Marine Biology
 \\University of California Santa Barbara\\Santa Barbara, CA 93106 USA\\
 dokamoto@sfu.ca\\
 
 (Present address: School of Resource and Environmental Management \& Hakai Network\\ Simon Fraser University, Burnaby, BC V5A 1S6, Canada)
}
\date{Updated Dec 11, 2015}
\maketitle
\begin{center}
\end{center}
\begin{flushleft}
\textbf{Citation}:  Okamoto, D.K. (in press). Competition among eggs shifts to cooperation along a sperm supply gradient in an external fertilizer.  \emph{The American Naturalist}. 

 \end{flushleft}

\section*{Abstract}

Competition among gametes for fertilization imposes strong selection.  For external fertilizers, this selective pressure extends to eggs for which spawning conditions can range from sperm limitation (competition among eggs) to sexual conflict (overabundance of competing sperm toxic to eggs).   Yet existing fertilization models ignore dynamics that can alter the functional nature of gamete interactions.  These factors include attraction of sperm to eggs, egg crowding effects or other nonlinearities in per capita rates of sperm-egg interaction.  Such processes potentially allow egg concentrations to drastically affect viable fertilization probabilities.   I experimentally tested whether such egg effects occur using the urchin \emph{Strongylocentrotus purpuratus} and parameterized a newly derived model of fertilization dynamics and existing models modified to include such interactions.   The experiments revealed that at low sperm concentrations, eggs compete for sperm while at high sperm concentrations eggs cooperatively reduce abnormal fertilization (a proxy for polyspermy).   I show that these observations are consistent with declines in the per capita rate at which sperm and eggs interact as eggs increase in density.   The results suggest a fitness trade-off of egg release during spawning:  as sperm range from scarce to superabundant, interactions among eggs transition from highly competitive to cooperative in terms of viable fertilization probabilities.

\section*{Introduction}

Fertilization of eggs tends to follow a fitness ``Goldilocks" principle.   Too few sperm results in low probability of fertilization while too many leads to polyspermy (multiple sperm entering one egg) and egg death \citep{franke2002,levitan2006,levitan2007,sewell2013,levitanbook2010}.     Because egg production is generally costly to females, balancing sperm-limitation versus sperm toxicity can confer a substantial fitness advantage \citep{levitan2004}.   Empirical and theoretical exploration of this trade-off in broadcast spawners has focused heavily on the role of sperm availability.  In contrast, the particular manner in which egg concentrations alter both fertilization and polyspermy remains less explored.   Specifically the per capita probability that sperm encounter eggs (per sperm, per egg) may vary with egg concentrations (i.e. a nonlinear functional response).  Such interactions are currently not considered and have the potential to dramatically alter expectations of both fertilization and polyspermy.  

For more than a century individuals have argued that egg concentrations likely alter sperm behavior and fertilization dynamics potential consequences for fitness \citep{lillie1915,loeb1914,mcclary1992, levitan1991,levitan2005,vogel1982,marshall2005, henshaw2014}.  Eggs are experimentally known to compete for sperm \citep{vogel1982,marshall2005} and total fertilization rates have been observed to shift downwards as sperm:egg ratios decrease \citep[e.g.][]{Benzie1994}.    Egg concentrations have the potential to impact fertilization maxima in two primary ways.  First, as egg concentrations increase and the sperm-egg ratio decreases, more sperm may be needed to maximize fertilization as more eggs are available to draw down the population of viable sperm [i.e. exploitative competition as inferred by \cite{vogel1982}, \cite{marshall2005} and \cite{Benzie1994}].   Second, the per capita rate of sperm-egg interactions (also called ``collision rates") may change with egg concentrations because of behavioral shifts due to attraction of sperm to eggs or interference among eggs (nonlinear interactions).  If the nature of sperm-egg interactions changes with the concentration of eggs then fitness of females may depend not only on sperm concentrations or sperm-egg ratios, but also on the rate of egg release and proximity to other females during spawning.   Moreover, such a scenario presents a possible shift in the nature of egg competition along a sperm gradient.  Specifically if higher egg concentrations lead to lower interactions with sperm per egg, this may exacerbate competition at low sperm concentrations but ameliorate polyspermy at high concentrations.  

Under existing models of broadcast spawning with purely random sperm-egg collisions, the per-capita rates of interaction are static \citep{vogel1982,levitan1991,bode2007,lauzon-guay2007}.   These models assume collision rates are unaffected by attraction of sperm to eggs, non-random search patterns or egg crowding effects.   Yet sperm can exhibit attraction \citep{kaupp2006,zimmer2011,evans2012}, structured search patterns \citep{farley2002} and chemical cues from eggs can induce behavioral changes in swimming direction \citep{guerrero2010} and swimming speed \citep{wood2007}.   These properties provide strong potential that per capita collision rates (no. per individual per second) are not random but decrease as egg concentrations increase.    

Empirically demonstrating the presence of interaction parameters that vary with density presents a serious challenge for many biological processes, including disease transmission \citep{mccallum2001} and predator-prey interactions \citep{cosner1999} in part because directly observing the interactions remains difficult.   Instead interaction rates are estimated from the product of interactions integrated over space and time (i.e. total number of people infected,  prey consumed or fertilized zygotes produced), yet this product often results from a complicated, unknown and highly variable series of processes.  Because of its mathematical simplicity, consistency in per capita interaction rates (a type I functional response) is commonly assumed.  Whether or not this assumption is valid can have significant theoretical and practical consequences.

Despite arguments in favor of the random sperm-egg collision model \citep[also described as ``mass action''; ][]{vogel1982}, sperm and eggs from free-spawning species have the potential to exhibit per capita collision rates that vary with egg concentration.   For example, sperm that are attracted to eggs may move towards eggs when flow dynamics allow such autonomous behavior \citep{zimmer2011,riffell2007}.   To illustrate how this behavior leads to nonlinear per capita rates of interaction, consider the following two circumstances: A)  eggs are abundant and densely concentrated or B) eggs are rare and sparsely concentrated.  Assume that sperm concentrations are initially identical between the two.  Sperm that are attracted to eggs will aggregate in higher density around individual eggs in case (B) than case (A).  This greater sperm density leads to higher overall interaction rates per individual in (A) than (B) when averaging over the area occupied by eggs.   Thus, as egg concentrations vary from dense to sparse, the per capita collision rate increases and is likely to saturate at some maximum (restricted perhaps by limitations to attraction and motility).  This is sensible given that chemoattractants can increase the ``target size'' of eggs \citep{jantzen2001}, which is directly related to the collision rate.  In contrast, purely random interactions (assumed by most models) yield identical per capita interaction rates between cases (A) and (B).  Other mechanisms beyond attraction may yield similar nonlinear collision functional responses, such as crowding (where increasing concentrations of eggs can inhibit search patterns of sperm, for example).   In this way, egg concentrations may impact fertilization rates by altering sperm behavior in addition to direct exploitative competition.  

If autonomous movement such as attraction is responsible for nonlinearities in per capita collision rates, then patterns in nature are likely to only exhibit such dynamics contextually.  Specifically, water velocity and shear stress can alter the effectiveness of attractive behavior because the capacity for autonomous motion of sperm diminishes \citep{zimmer2011,riffell2007}.   In this case, laboratory systems are unlikely to relate well to nature and direct comparisons across taxa from lab experiments may not be possible.  Bias may result from using or comparing highly constrained or conditional subsets of the parameter space.   Thus, from both a theoretical and practical standpoint it is important to understand the extent to which fertilization is governed by per capita interaction rates that vary by egg concentration.  

In this study I evaluate whether egg concentrations impact fertilization maxima, and explicitly test for random versus nonlinear sperm-egg collision rates and whether the nature of interactions among eggs shifts from competition to facilitation along the gradient of sperm availability.  To achieve these goals I developed a new dynamic model that expands upon existing models and then conducted a laboratory fertilization experiment using purple sea urchin (\emph{Strongylocentrotus purpuratus}) eggs and sperm.  Finally I parameterized and compared the performance of models that included random or nonlinear per-capita collision parameters by utilizing several different basic model forms and my empirical observations of fertilization.   

\section*{Methods}
In order to evaluate how the trade-off in sperm concentration is affected by egg concentrations I conducted multiple trials of a laboratory experiment that vary concentrations of eggs and sperm.    I used two analytic approaches to generate inference about the nature of sperm-egg interactions.  First I used statistical models to test hypotheses about the data and second used mechanistic models to estimate to what extent competition among eggs and behavioral shifts in sperm-egg collision rates alter fertilization rates.   For the former approach I estimating generalized linear mixed effects models and tested whether sperm and egg concentrations influence successful and abnormal fertilization.  For the latter I used three distinct model forms (including a newly generated dynamic model) to compare constant vs nonlinear sperm-egg collision rates using a Bayesian hierarchical framework.  This method is useful for comparing hypothetical dynamic processes rather than just emergent properties of the system.  

\subsection*{Expanded dynamic fertilization models}
\subsubsection*{Existing models}
Existing fertilization kinetics models, and interpretations thereof, assume purely random interactions between sperm and eggs.  This means the instantaneous per capita rate of sperm-egg interactions (number of interactions per individual) are independent of sperm and egg concentrations.  Specifically, the model of \cite{vogel1982} and extensions to include effects of polyspermy by \cite{styan1998} and \cite{millar2003} are based upon a system of equations that describe the loss of viable sperm ($S$) and the reduction in unfertilized eggs (${E}_{U}$) over time.  These models assume: (1) sperm ($S$) collide with a fixed population of eggs (${E}_{T}= $ fertilized plus unfertilized eggs) at a constant, per capita rate ${\beta}$, (2) a fixed proportion of sperm collisions result in fertilization ($\gamma$), and (3) a sperm collision with an egg (successful or unsuccessful) renders that sperm inert.   Such dynamics are given by eqns \eqref{eq:DS1} and \eqref{eq:DE1}: 

\begin{subequations}
\begin{align}
\frac{dS}{dt} &=-{\beta} S(t){E}_{T}\label{eq:DS1}\\
\frac{d{E}_{U}}{dt} &=-\gamma{\beta} S(t){E}_{U}\label{eq:DE1}
\end{align}
\end{subequations}
This system of differential equations has solution given by eqns. \eqref{eq:S1} and \eqref{eq:E1}:

\begin{subequations}
\begin{align}
S(t)&={S}_{0}\exp\left(-t\beta {E}_{T}\right)\label{eq:S1}\\
{E}_{U}(t)&={E}_{T}\exp\left(-\gamma \frac{{S}_{0}}{{E}_{T}} \left(1-\exp\left(-t\beta {E}_{T}\right)\right)\right)\label{eq:E1}
\end{align}
\end{subequations}
\cite{styan1998} and subsequently  \cite{millar2003} used this solution to generate an equation for the number of eggs that are fertilized and viable (${E}_{M}$, monospermic zygotes); hereafter their models are referred to as the Styan and MA models (see Appendix A for the explicit model forms).     These models explicitly describe the hypothetical process by which eggs mount a defensive polyspermy block following the first successful sperm to penetrate the jelly layer.  They consider the block to be a step function in that some time period after the first invader eggs are no longer penetrable, but are fully susceptible in the interim to a second fertilizer.   

There are several potential drawbacks to using the \cite{vogel1982} and related formulations to test for nonlinearities in per capita collision rates.  First, there is no explicit term for sperm degradation over time.  Instead \cite{vogel1982} suggest substituting a sperm ``half-life" ($\tau$) for the duration of sperm-egg ``contact time" $t$ if the $t>\tau$.     This assumes that the asymptotic dynamics are equivalent to a step function where there is no degradation until contact time reaches $\tau$, and thereafter 100\% of the sperm are not viable.  This approach has the capacity to produce different results than a model including a dynamic process of sperm viability decay.  Moreover, the \cite{vogel1982} method of estimating $\tau$ is biased towards higher values as sperm concentrations increase (see Appendix A).  Second, the derivation of the models do not explicitly incorporate additional compartments in the differential equations, which in this case include a) fertilized eggs vulnerable to polyspermy, b) polyspermic eggs, and c) monospermic zygotes invulnerable to a second sperm.  These are potentially valuable components for applying the model beyond a constrained laboratory study to an open, dynamical system.  Thus, while such models perform well in explaining laboratory data, they potentially introduce bias due to the sperm decay issues and are often not applicable to more complex or dynamic systems. 

Below I explain an expanded system of differential equations that has the flexibility to address these issues.  The value of such a system of equations is that the model can be expanded to incorporate a dynamic sperm-egg behavioral functional response, sperm degradation function or to a system of partial differential equations (for time and space) in order to account for advection or diffusion of sperm and eggs or introduction of fresh sperm and eggs as additional individuals spawn in space and time (see the appendix D for an animated example).  In addition, such systems can be expanded to include among-pair variance in parameters; for example one can include pairwise gamete compatibility \citep{styan2008} by giving gametes from each parent a unique set of equations and parameter values in the system with parameters estimated using hierarchical models (see below).   Moreover the system above can easily be applied to laboratory settings for testing of hypotheses concerning fertilization dynamics.

\subsubsection*{Updated dynamic model}
I built upon the existing set of differential equations constructed by \cite{vogel1982} to generate an updated, dynamic model.  This model includes a fully compartmentalized system of differential equations.  I thereafter expand this model as well as that of \cite{styan1998} and \cite{millar2003} to allow for nonlinear per capita collision rates for each of these three model forms.  In the case of the new dynamic model,  sperm ($S$) decay (per capita) at a natural rate $r$ after release.  This is similar to a recent model \citep[model 2.7 in][]{lehtonen2015} in that the mortality of sperm is accounted for explicitly, but eggs are considered indefinitely viable (though this can easily be extended to consider a natural decay function in egg viability). Like eq. \eqref{eq:DS1}, sperm collide with eggs ($E_T$) at a constant rate ${\beta}$ (nonlinearities are accounted for later).  The overall loss rate of sperm is given by eq. \eqref{eq:1}:

\begin{subequations}
\begin{equation}\label{eq:1}
\frac{dS}{dt} =-{\beta} S(t){E}_{T}(t)-rS(t)\end{equation}
Unfertilized eggs (${E}_{U}$) are fertilized at a rate given by eq. \eqref{eq:2}:
\begin{equation}\label{eq:2}
\frac { { { dE }_{U} } } {dt} =-\gamma {\beta}  S(t){E}_{U}(t)
\end{equation}

but newly fertilized eggs (${E}_{V}$) are still vulnerable to a second fertilizer.   If these eggs induce a polyspermy block at rate $\delta$ and are fertilized by a second sperm at the same rate as unfertilized eggs, then the rate of change of vulnerable eggs is given by eq. \eqref{eq:3}:

\begin{equation}\label{eq:3}
\frac{{{dE}_{V}}}{dt} =\gamma {\beta}  S(t){E}_{U}(t)-\delta{E}_{V}(t)-\gamma {\beta}  S(t){E}_{V}(t)
\end{equation}

Clearly this is a simplistic representation of polyspermy block dynamics, but I use this representation for simplicity.  Eggs that successfully induce a polyspermy block without being fertilized by a second sperm (${E}_{M}$) accumulate at the rate given by eq. \eqref{eq:4}:

\begin{equation}
\label{eq:4}
\frac{{{dE}_{M}}}{dt} =\delta{E}_{V}(t)
\end{equation}

while eggs that become fertilized by a second sperm (${E}_{P}$) accumulate at the rate given by eq. \eqref{eq:5}:
\begin{equation}\label{eq:5}
\frac{{{dE}_{P}}}{dt} =\gamma {\beta}  S(t){E}_{V}(t)
\end{equation}
\end{subequations}

For parameter descriptions see Table \ref{table:parametertable}.  For a simple example of how this model can be expanded to incorporate spatial dimensions, specifically sperm diffusion and competition among upstream and downstream eggs see Appendix D.

\subsubsection*{Application to closed laboratory environments}
If the system is closed (i.e. no advection) and fresh, virgin sperm and eggs are introduced all at once to the system at time $t=0$, then the subsystem which includes equations \eqref{eq:1}, \eqref{eq:2}, and \eqref{eq:3} can be integrated analytically with respect to $t$ (and does not depend on equations \eqref{eq:4} or \eqref{eq:5}).  In this case, ${E}_{T}$(t) becomes the constant ${E}_{T}$ (initial number of eggs), all eggs at time $t\approx 0$ are unfertilized (i.e. ${E}_{U}(t=0)={E}_{T}$) and let $S(t=0)=S_0$.   The solutions in such a circumstance are given by eq. \eqref{eq:4a}-\eqref{eq:4c}:

\begin{subequations}
\begin{align}
{S}(t) & ={S}_{0} \exp \left(t \left(-\beta _0 {E}_{T}-r\right)\right)\label{eq:4a}\\
\notag\\
{E}_{U}(t) & ={E}_{T} \exp \left(-\frac{\beta _0 S_0 \gamma }{\beta _0 {E}_{T}+r}-\frac{\beta _0 {S}_{0} \gamma  e^{-t \left(\beta _0 {E}_{T}+r\right)}}{-\beta _0 {E}_{T}-r}\right)\label{eq:4b}\\
\notag\\
{E}_{V}(t) & =\frac{{\beta} {E}_{T} {S}_{0} \gamma \left(e^{t ({\beta} {E}_{T}-\delta+r)}-1\right) }{\exp \left(t ({\beta} {E}_{T}-\delta+r)+\frac{{\beta} {S}_{0} \gamma e^{t (-({\beta} {E}_{T}+r))}}{-{\beta} {E}_{T}-r}+\frac{{\beta} {S}_{0} \gamma}{{\beta} {E}_{T}+r}+\delta t\right)\left({\beta} {E}_{T}-\delta+r \right)}\label{eq:4c}
\end{align}

\begin{table*}[t]
\caption[Dynamic model parameters and state-varibles]{Description of parameters and state-variables in the system of differential equations described by eqns. \eqref{eq:1}-\eqref{eq:5}.}
\begin{small}
\label{table:parametertable}
 \begin{threeparttable}
 \centering
\begin{tabular*}{\textwidth}{c @{\extracolsep{\fill}} lc}
\hline
Parameter & Description \\
\hline
${\beta}$ & sperm collision rate\\
$\gamma$ & egg selectivity\\
$\delta$ & per capita polyspermy block rate\\
$r$ & viable sperm decay rate\\
$S_0$ & initial sperm concentration\\
${E}_{T}$\mtnote{$\dagger$} & total number of eggs in the system\\
\hline
State Variable & Description \\
\hline
${E}_{U}$ & unfertilized eggs in the system\\
${E}_{V}$ & fertilized eggs vulnerable to polyspermy\\
${E}_{M}$ & monozygotic eggs invulnerable to polyspermy\\
${E}_{P}$ & eggs fertilized by multiple sperm (polyspermy)\\
${S}$ & viable sperm in the system\\
\hline
\end{tabular*}
\begin{tablenotes}
\footnotesize
            \item[$\dagger$] In a closed environment $E_T$ is fixed whereas in an open system where eggs are advected $E_T$ can be represented by a state variable and additional equations.
        \end{tablenotes}  
 \end{threeparttable}
 \end{small}
 \end{table*}
 
 The number of eggs successfully fertilized by a single sperm that are invulnerable to a second sperm (the integral of eq. \eqref{eq:4}) cannot be generated analytically.   However, because eq. \eqref{eq:4} is solely a function of $\delta$ and ${E}_{V}(t)$, its solution can be expressed as given by eq. \eqref{eq:4d}:
 \begin{equation}\label{eq:4d}
 {E}_{M}(t)=\delta \int_{ 0 }^{ t }{E}_{V}(t)dt\end{equation}
 \end{subequations}
where ${E}_{V}(t)$ is given by eq. \eqref{eq:4c}.  This is now a single differential equation and the integral can quickly and accurately be approximated numerically using simple numerical integration methods (see: \emph{Parameter estimation for the dynamic model and other fertilization models} below).  

\subsubsection*{Addition of non-random sperm-egg interactions to fertilization models}
Non-random sperm-egg interactions can be incorporated into models of fertilization in a variety of ways.  The simplest manner of incorporating such dynamics is by allowing the rate at which sperm attack eggs (${\beta}$) to vary with egg concentration.   One justification for such an approach is the fact that chomoattractants can act to increase the ``target size'' of eggs.  In this case I allow unique collision rates (${{\beta}}_{i}$) for each egg concentration treatment which is valuable given there is no known functional relationship that drives non-random sperm egg interactions.  One could easily employ a Type II functional response \citep{holling1959,cosner1999} on the per capita collision parameter, but without extensive experimentation there is no way to evaluate what shape is biologically appropriate \emph{a priori}.  

\subsection*{Laboratory experiments}
I conducted eight laboratory trials of an experiment with the purple sea urchin \emph{Strongylocentrotus purpuratus} between February 23 and March 8, 2014 during their spawning season.   Each trial included a factorially crossed gradient of four egg concentrations ($\approx$ 1, 0.25, 0.0625 \& 0.0126 $\mu l^{-1}$) with six sperm concentrations ($\approx$ $10^{-3}$, $10^{-4}$, $10^{-5}$, $10^{-6}$, $10^{-7}$ \& $10^{-8}$ $\mu l^{-1}$) using gametes from one male and one female per trial.  The experiment took place in a climate-controlled room at 13\degree C, and each experimental unit consisted of a 20 ml vial with 10 ml seawater.  For each trial I first added 8 ml seawater to the 24 experimental vials, followed by 1 ml appropriate egg solution and finally providing each vial with 1ml freshly diluted sperm solution.  Vials were gently agitated by hand and placed on an orbital shaker table at $\approx1.5$ revolutions per second and left for 120 minutes to allow fertilization and first cleavage to occur.  Following the 120 minute incubation period, vials were agitated by hand and emptied into shallow petri dishes for examination under an inverted compound microscope where 100 random, undamaged eggs were scored for fertilization, cleavage and polyspermy.  Fertilization criteria included the presence of a raised vitelline membrane or cell division.  Cleavage criteria included only cells that divided normally (radial division) at least once.  While positive recognition of polyspermy requires counting of pronuclei that entered the cell \citep{franke2002,levitan2004}, abnormal cell cleavage associated with polyspermy is often used as a proxy for polyspermy, which I hereafter refer to as ``abnormal fertilization".  While possible that abnormal fertilization arises from other mechanisms related to sperm concentrations, this is unlikely to differ among egg concentrations used here.  Moreover previous work with various urchin species show strong evidence for polyspermy at high sperm concentrations inferred from direct measurement \citep{levitan2004, franke2002} and abnormal development consistent with polyspermy \citep{styan1998,styan2008,levitan2006,levitan2007,sewell2013,levitanbook2010}.  Thus I inferred polyspermy from tetrahedral or further abnormal division of the cell.  Very few fertilized eggs did not divide after 120 minutes and were scored as viable; as such, abnormal fertilization may be underestimated in those few cases. 
 
Urchins were collected in collaboration with the Santa Barbara Coastal Long Term Ecological Research (SBC LTER) program at a depth of approximately 7 m below mean low water at the Mohawk Reef (119.7296 W,  34.3941 N) near Santa Barbara, California in late February, 2014.  I maintained urchins in flow through seawater tables until needed (generally 3-5 days).  To obtain eggs, I induced females to spawn by injecting 1 ml 0.55M KCl adjacent to the Aristotle's lantern, gently agitated, and placed them upright into a small container with seawater.   Once a female began profusely extruding eggs, I extracted 1 ml of concentrated egg material in 200 $\mu$l batches directly from the gonopores and placed into 50 ml seawater.     I then added appropriate dilutions of this solution to each vial such that vials had final concentrations of approximately 1, 0.25, 0.0625, or 0.0126 eggs $\mu l^{-1}$.   To obtain sperm, I injected males with KCl in the same manner as females but extracted 100 $\mu$l ``dry'' sperm directly from the gonopore without submerging the animal.   Sperm were immediately diluted $100 \times$ in seawater, directly followed by six serial $10 \times$ dilutions.  Sperm solutions were then added in 1 ml aliquots to the vials prepared with eggs (final concentrations $ \approx 10^{-3}, 10^{-4}, 10^{-5}, 10^{-6}, 10^{-7} \text{ }\&\text{ } 10^{-8}  \text{ }\mu l^{-1}$).   Actual sperm concentrations in vials varied slightly because of small variability in sperm counts  (see Figure 1 for range).  Note that the highest egg concentration was chosen such that eggs would not form a densely packed layer on the bottom of the vial.  Specifically, at 1 egg $\mu l^{-1}$, even if eggs were to completely sink to the bottom of the vial they would cover approximately 15\% of the vial bottom surface area, assuming an average 0.084 mm egg diameter \citep{levitan1993} and an internal vial diameter of 22 mm.

To estimate sperm concentrations used in each trial, I preserved the $10^{-3}$ $\mu l^{-1}$ sperm solution with 2\% buffered formalin and conducted five replicate sperm cell counts on a hemocytometer.   I estimated egg concentrations for each trial by counting the number of eggs in five replicate 100 $\mu$l subsamples (agitated and homogeneous) using an inverted microscope.    

Data from the experiments are available from the LTER Network Data Portal \citep[http://dx.doi.org/10.6073/pasta/bf95a6c10ed0fa047f5b421492f2fa33, ][]{okamotodata}. 

\subsection*{Linear statistical analysis}
I first analyzed data using generalized linear mixed effects models (GLMMs) with a binomial likelihood and a logit link to evaluate whether egg concentrations in addition to sperm concentrations influence rates of fertilization and polyspermy.  I conducted separate analyses on the proportion of eggs that were fertilized and the proportion of eggs with polyspermy.  In each model, I imposed a random intercept for individual-male female combination (i.e. a random effect for trial).  Because the data exhibited overdispersion (greater error variance than expected under a simple binomial likelihood) I added a random, normal error error to the logit-scaled model probabilities.   Covariates for fertilization included a third-order polynomial of the log of sperm concentration and log of egg concentration in the full model due to the shape of the response.   For polyspermy I used log of sperm concentration and log of egg concentration in the full model.  I tested hypotheses of individual effects of eggs and sperm on fertilization using likelihood ratio tests.   I estimated GLMMs using the R package lme4 \citep{lme4}.  I employed these statistical models to evaluate whether fertilization rates responded to sperm and/or egg concentrations.   

\subsection*{Parameter estimation for the dynamic model and other fertilization models}

In contrast to the statistical models described above used for hypothesis testing, I employed the mechanistic models to estimate 1) whether per capita collision rates are non-random and instead vary by egg concentration and 2) the consequences of such changes on expected fertilization dynamics.   Thus for each model form I estimated models that included unique sperm collision rates (${\beta}$) for each mean egg concentration.  I estimated all models within a Bayesian hierarchical framework via the no-U-turn sampler (NUTS) variant of Hamiltonian Monte Carlo (HMC) \citep{hoffman2014} in Stan via the rstan package\citep{stansoftware,rstansoftware} using R \citep{Rsoftware}.   The hierarchical framework is essential to account for potential among-pair variability in dynamics while also estimating overall means for those parameters.  I included hierarchical effects for sperm collision rates (${\beta}$), egg selectivity (${\gamma}$) and the polyspermy block rate (${\delta}$) to account for potential variability in rates among pairs   I assume no among-pair variability in other parameters.   For details of the hierarchical structure see appendix C.   I employed a Bayesian approach because it integrates inference over the plausible range of parameter values while point estimation procedures (e.g. maximum likelihood or nonlinear least squares) in fertilization kinetics models can struggle with consistency and may not produce unique parameter solutions \citep{styan2000}.

Because there is no closed form solution for $E_M(t)$, there is also no closed form solution for the likelihood of the stochastic process.  Thus, for simplicity and consistency across models that are derived differently (the Styan, MA and dynamic models), I used a beta-binomial likelihood for all models.  The beta-binomial is a mixture distribution that relaxes the constrained variance of the binomial by including a variance dispersion parameter $\lambda$ for the mean, which when $\lambda \rightarrow \infty$ the distribution converges to a standard binomial.   
\begin{figure*}[t]
\begin{center}
\includegraphics[width= 100mm]{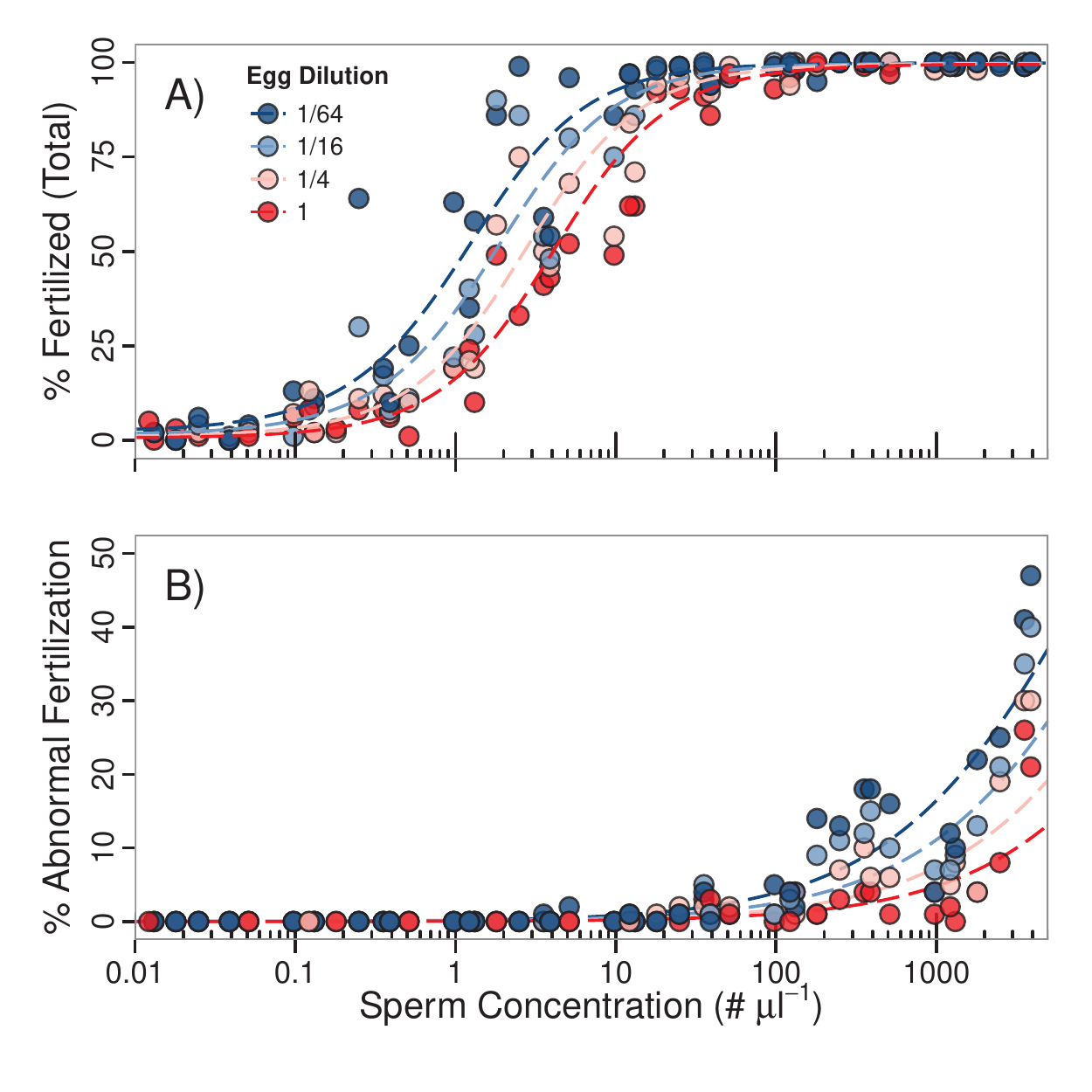}
\end{center}
\caption[Experimental observations of fertilization and polyspermy]{Percent total egg fertilization (A) and percent abnormal fertilization (a proxy for polyspermy) (B) in experimental treatments.   Broken lines represent GLMM model fixed effects (see Table 2);  deviation from the fixed effect consists of both error within and among individual pairs.  Egg dilution represents the mean concentration ($\# {\mu l}^{-1}$) of eggs.  Colors indicate different egg concentrations, with blue representing lower concentrations.  } 
\label{fig:F1}
\end{figure*}

 All parameters are described, along with their priors, in Table \ref{table:parametertable}.  For each model form (the Styan and MA models and the dynamic model) all parameters were given uniform priors over the boundaries of their realistic range except $\lambda$ which was given a Pareto prior with minimum 1 and shape 1.5 \citep{gelman2013}.   Sampling included 1500 total iterations from three chains, each of which included a 500 iteration burn in period which was sufficient to produce convergence \citep{gelman2013}.  Because  \eqref{eq:4d} does not have an explicit form, each iteration in the MCMC chain of the dynamic model included generating the numerical solutions of the expected value for each observation given parameter values using Gauss-Legendre quadrature rules.    A constant sperm half-life was used for the Styan and MA models (though results are qualitatively insensitive to whether a constant value or a sperm concentration-dependent equation provided by \cite{levitan1993} is used; see Appendix B for estimation of the sperm decay rate $r$ for the dynamic model, as well as for estimation and justification for using a constant sperm "half-life" for the Styan and MA models).

Finally, to evaluate whether the nonlinear (non-random) collision rates improve predictive accuracy, for each model form I compared the nonlinear collision form with the null (random collision) using WAIC (widely applicable information criterion, \cite{watanabe2010}).  WAIC is a fully Bayesian information criterion that is the log pointwise posterior predictive density plus a correction for effective number of parameters \citep{gelman2013}.  I use the convention of \cite{gelman2013} by multiplying the WAIC of \cite{watanabe2010} by -2 to put it on the deviance scale (thus on the same scale as AIC, BIC or DIC).   On this scale, lower values of WAIC are favored.  

\section*{Results}
There was a clear trade-off between maximizing total fertilization and minimizing polyspermy in terms of both egg and sperm concentration.  Egg concentration negatively affected the rate of fertilization, requiring much more sperm to achieve similar fertilization rates with higher egg concentrations (Figure 1A, Table \ref{table:stat.table}).  The densest egg concentrations (1 eggs $\mu l^{-1}$) required roughly five times more sperm than sparsest egg concentrations ($\frac{1}{64}$ eggs $\mu l^{-1}$) to achieve the same fertilization rate (Figure 1A) with intermediate egg concentrations showing intermediate sperm requirements.  In contrast rates of polyspermy decreased with egg concentration, where low concentrations of eggs resulted in much higher rates of polyspermy (Figure 1B, Table \ref{table:stat.table}).  Sparse egg concentrations increased rates of polyspermy approximately threefold at the highest sperm concentrations compared to the most concentrated egg samples.  

\begin{table*}[t]
\caption[GLMM model coefficients and tests]{Parameter estimates, confidence intervals and likelihood ratio tests for generalized linear mixed effects models (with binomial likelihood and logit link).  Predictions for fixed effects are represented in Figure 1.  Biological model results are shown in Table \ref{table:param.values}.   All models include a random effect of trial pair (i.e., intercepts vary by individual male-female pairs) listed as $\sigma_{Intercept}$ and a logit-scale error variance (incorporated to account for overdispersion) listed as $\sigma_{Error}$.  Predictor variables (sperm and egg concentration) were $\log_{10}$ transformed in the analyses.  A 3rd order polynomial for sperm concentration was included \emph{a-priori} and the significance of the combined polynomial was tested (i.e., no test for separate polynomial effects).  Note that $\sigma_\text{Error}$ is the imposed normal error variance added to the logit scale linear predictors to account for overdispersion, while the binomial error sd is not an estimated parameter.}
\begin{small}
\begin{center}
\begin{tabular*}{\textwidth}{c @{\extracolsep{\fill}} l|rcccc}
  \toprule
Response & Parameter & Mean & 95\% CI & $\chi^{2}$ & df & P-value \\ 
  \midrule
 & $\log_{10}$(Eggs) & -0.81 & -1.02 : -0.61 & 544.5 & 1 & $<0.0001$ \\ 
  Total & $\log_{10}$(Sperm) [ 1\degree polynom. ] & 48.13 & 45.48 : 51.04 & 2258.4 & 3 & $<0.0001$ \\ 
  Fertilization & $\log_{10}$(Sperm) [ 2\degree polynom. ] & -0.70 & -3.03 : 1.74 &  &  &  \\ 
  (Fig 1A) & $\log_{10}$(Sperm) [ 3\degree polynom. ] & -6.22 & -8.44 : -3.98 &  &  &  \\ 
   & Intercept & 0.54 & 0.13 : 0.95 &  &  &  \\ 
   & $\sigma_{\text{Intercept}}$ & 0.43 & 0.24 : 0.83 &  &  &  \\ 
   & $\sigma_{\text{Error}}$ & 0.69 & 0.56 : 0.84 &  &  &  \\ 
   \\ 
 & $\log_{10}$(Eggs) & -0.75 & -0.93 : -0.59 & 48.5 & 1 & $<0.0001$ \\ 
  Polyspermy & $\log_{10}$(Sperm) & 1.58 & 1.43 : 1.75 & 297.4 & 1 & $<0.0001$ \\ 
  (Fig 1B) & Intercept & -7.71 & -8.38 : -7.10 &  &  &  \\ 
   & $\sigma_{\text{Intercept}}$ & 0.51 & 0.31 : 0.95 &  &  &  \\ 
   & $\sigma_{\text{Error}}$ & 0.22 & 0.00 : 0.42 &  &  &  \\ 
   \bottomrule
\end{tabular*}
\end{center}
\end{small}
\label{table:stat.table}
\end{table*}

\subsection*{Fertilization models}

In all model forms [eq. \eqref{eq:4d} or that of \cite{styan1998} or \cite{millar2003}] sperm-egg collision rates declined substantially as egg concentration increased.    Compared to the lowest egg concentration ($\frac{1}{64}$ eggs $\mu {l}^{-1}$), the sperm collision rate declined by an estimated 44\%, 59\% and 74\% at  $\frac{1}{16}$, $\frac{1}{4}$, and $1$ eggs ${\mu l}^{-1}$ (Figure 2), respectively in the dynamic model (given as a proportion of the collision rate at $\frac{1}{64}$ eggs ${\mu l}^{-1}$ in Table \ref{table:param.values}).   None of the 95\% credible sets for the decreases in collision rates overlap with 0.   A similar decrease in collision rate was found for all three model forms (Figure 2).  Predictions from the nonlinear collision model forms were consistent with the results of the experiment, reproducing the observed separation in fertilization curves between egg treatments (Figure 3).  The model including non-random collision rates performed better with respect to WAIC than the random collision model (WAIC of 1304.0 versus 1329.2, respectively where lower values are preferred).  
\begin{figure*}[t]
\begin{center}
\includegraphics[width= 100mm]{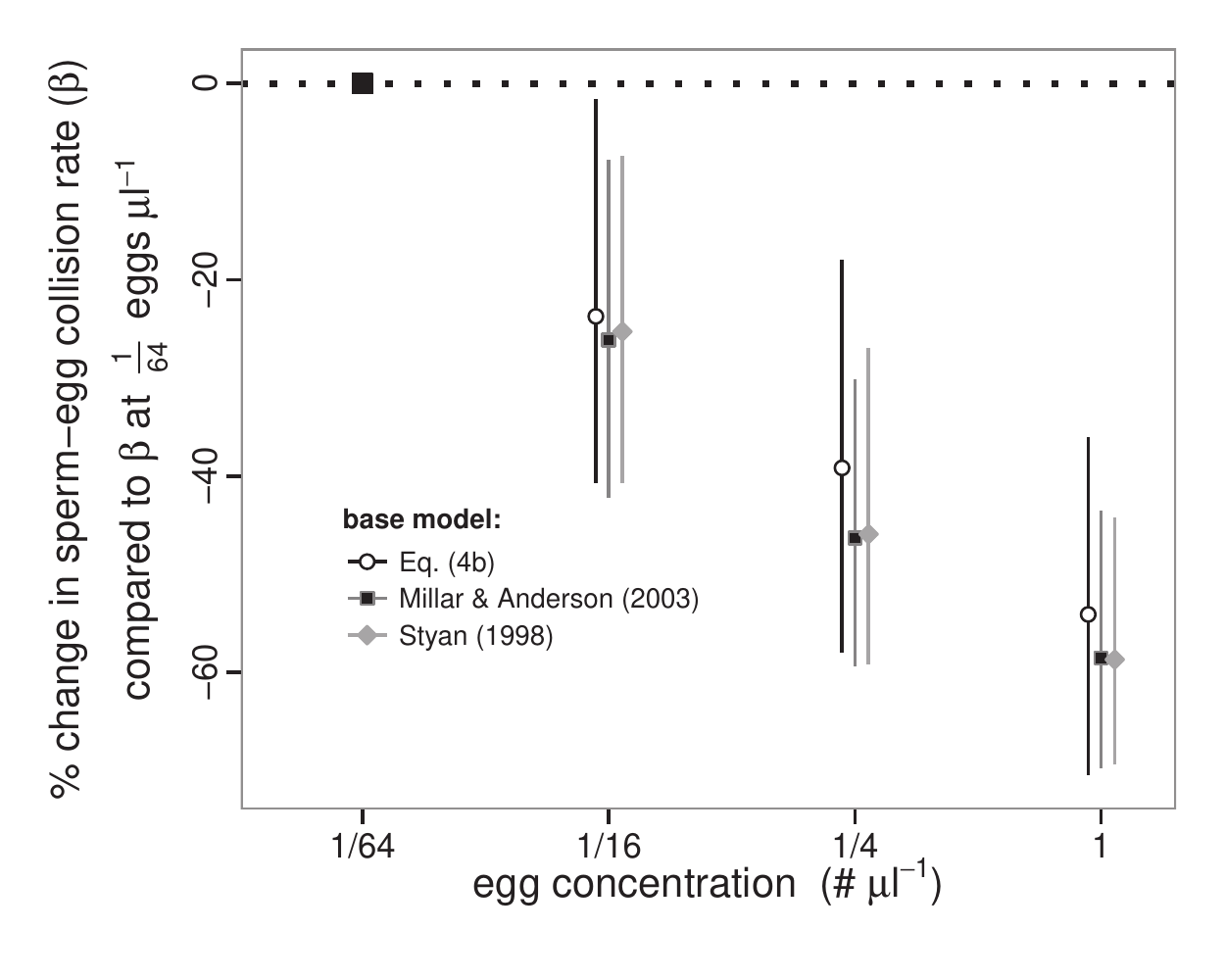}
\end{center}
\caption[Change in sperm-egg collision rate]{Estimated \% change in collision rate (${\beta}$) at each mean egg concentration (egg concentration among pairs varied slightly) from the collision rate at 1/64 eggs $  {\mu l}^{-1}$ for each model.  Each unique collision rate is estimated as the collision rate at 1/64 eggs $ {\mu l}^{-1}$ times the value shown.  Egg dilution represents the mean concentration ($ {\mu l}^{-1}$).}
\label{fig:F2}
\end{figure*}

In terms of model expectations, the largest effect of non-random collision rates was in the rate of abnormal fertilization.   Predictions when including different per capita collision rates at each egg concentration were consistent with the results of the experiment (Figure 4 C). When collision rates were held constant the predictions cannot explain the separation among egg treatments at high sperm concentration (Figure 4 D) [note that this observation depends on the circumstances as particular combinations of parameters that not supported here may lead to greater egg concentrations right-shifting abnormal fertilization under different scenarios].  In other words, the effects of egg concentrations on abnormal fertilization in this case cannot be explained without including nonlinear per capita collision rates.  

In contrast, the differences in fertilization curves among egg concentrations (Figure 1A) can be explained in part by exploitative competition for sperm by eggs.  Specifically, differences in viable fertilization at low sperm concentrations are still expected without substantial variation in collision rates (Figure 4 A,B).  This is because when sperm are rare, egg concentrations have a greater capacity for sperm depletion and the impacts of variable collision rates less pronounced.    The reduction in fertilization rates under the random collision assumption occur solely a result of exploitative competition for sperm (i.e. more eggs use up more sperm, leaving fewer sperm collisions per egg on average).  In contrast, both interference and exploitative competition explain the phenomenon in the non-random collision model.  In this case, more eggs exist to attract sperm reducing the effective concentration around each egg (interference) \emph{and} more eggs use up more sperm.    

The observed pattern in Figure 1A can plausibly be explained via exploitation of sperm by eggs alone (i.e competition), while the observed divergence in Figure 1B requires nonlinear per capita collision rates to yield the combined curve shown in Figure 3.    The models and the data illustrate that increasing egg concentrations effectively shift the fertilization curve, thereby shifting the region of optimal sperm concentrations (Figure 3).  

In all models there was considerable uncertainty in the parameter estimate for egg selectivity ($\gamma$, Table \ref{table:param.values}).  This means that parameter estimates are averaged over the realistic range for this parameter.  Despite such uncertainty the models still suggest collision rates decline significantly as egg concentrations increase.   Thus, despite the lack of information about $\gamma$, there is sufficient information to support density dependent per capita collision rates over the null model (random collision rates).
\begin{table*}[t]
\centering
\caption{Table of parameters, description, estimated posterior mean, 95\% credible set, and the prior for the dynamic model with unique collision rates (U = uniform prior).  The full table is shown in Appendix A.   }
\label{table:param.values}
\begin{small}
 \begin{threeparttable}
\begin{tabular*}{\textwidth}{c @{\extracolsep{\fill}} cccc}

\hline
Parameter & Description & Mean & 95\% Credible Set & Prior\\
\hline
${{\beta}}_{\frac{1}{64}}$& median\mtnote{$\dagger$} collision rate ($\frac{1}{64}$)  & 2.59$\times {10}^{-3}$ & (0.99 : 7.62)$\times {10}^{-3}$ & U[1$\times{10}^{-5} : 0.1]$\\
${{\beta}}_{\frac{1}{16}}/{{\beta}}_{\frac{1}{64}}$ & collision rate ratio & 0.74 & 0.56 : 0.98 & U$[0.01 : 2.00]$\\
${{\beta}}_{\frac{1}{4}}/{{\beta}}_{\frac{1}{64}}$ & collision rate ratio & 0.59 & 0.42 : 0.84 & U$[0.01 : 2.00]$\\
${{\beta}}_{1}/{{\beta}}_{\frac{1}{64}}$ & collision rate ratio & 0.45 & 0.30 : 0.64 & U$[0.01 : 2.00]$\\
${\gamma}$ & mean egg selectively & 0.081 & 0.075 : 0.129  & U[1$\times{10}^{-5} : 0.15]$\\
${\delta}$& median\mtnote{$\dagger$} polypsermy block rate  & 1.84 & 1.00 : 3.26 & U[1$\times{10}^{-5}: 20.0]$\\
$\lambda$ & total count ($\text{dispersion}^{-1}$) & 31.1 & 23.8 : 39.4 & Pareto$[1,1.5]$\\
\hline
\end{tabular*}
\begin{tablenotes}
\footnotesize
            \item[$\dagger$] Among-pair median values are reported because the among-pair hierarchical distribution is assumed to be lognormal and thus the exponentiated posterior mean equals the median.
        \end{tablenotes}  
 \end{threeparttable}
 \end{small}
\end{table*}

Results are robust to the sperm ``half-life" that is used in the \cite{styan1998} or \cite{millar2003} models.  This occurs because using a longer or shorter half-life generally results in an inversely proportional change in the baseline collision rate because in all cases for these two models the two are multiplied together.  Thus, while exact estimates of the collision rates differ, the estimated change in collision rate with egg concentration does not.    

\section*{Discussion}
A rich body of evolutionary and ecological work is based upon the dynamics of sperm limitation in fertilization across a host of taxa \citep{levitan1998, levitan1996,marshall2007,bode2007, yund2000, levitanbook2010, levitan1993, podolsky1996, parker2014}.  Yet current models of external fertilization assume random collision rates between sperm and eggs; moreover empirical work tends to ignore the impact of egg concentrations \citep[but see ][]{vogel1982,levitan1991, marshall2005}.   If per capita sperm-egg collision rates are instead dependent upon egg concentrations this complicates the basic dynamics governing fertilization success and may provide additional targets of selective pressure.  Here I show that 1) controlled laboratory experiments produced observations consistent with non-random sperm-egg collision rates, and 2) models that explicitly include such dynamics are required to capture the observed behavior at high sperm concentrations.  These findings illustrate that changes in egg concentrations can cause considerable shifts in the range of sperm concentrations required for maximizing rates of fertilization.   
\begin{figure*}[t]
\begin{center}
\includegraphics[width= 100mm]{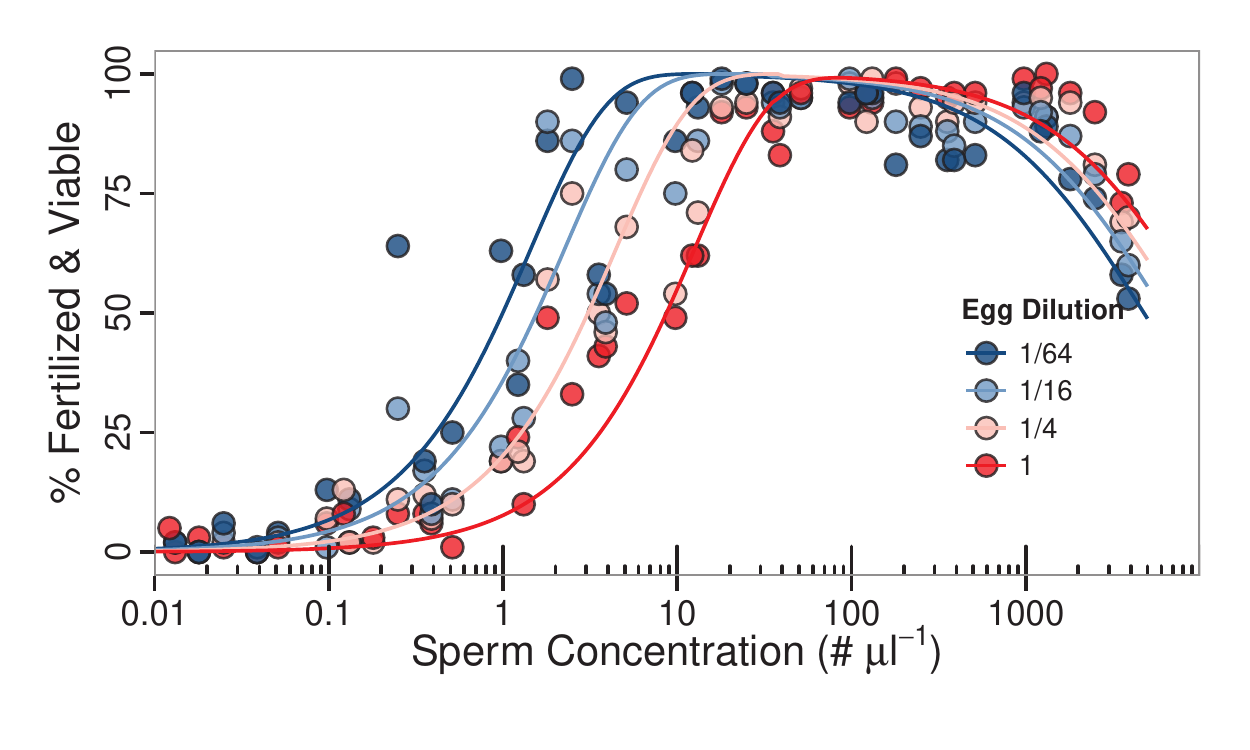}
\end{center}
\caption[Predictions under the full dynamic model]{Viable fertilization rate (percent total fertilization minus abnormal fertilization [a proxy for polyspermy]) in experimental treatments.  Lines represent the hierarchical mean from the dynamic model with nonlinear sperm-egg interactions (different collision rates by egg dilution). Egg dilution represents the mean concentration ($\# {\mu l}^{-1}$).  Models with purely random sperm-egg interactions do not capture the separation by egg dilution in rates of abnormal fertilization at high sperm concentrations.  Colors indicate different egg concentrations, with blue representing lower concentrations.}
\label{fig:F3}
\end{figure*}

These shifts appear driven by two simultaneously operating factors controlled by local egg concentrations.  First, a greater number of eggs results in quicker depletion of the sperm and, second, higher local concentrations of eggs depress the per-capita sperm-egg collision rate.   The consequences of these processes are not trivial given the importance of successful fertilization in maximizing lifetime fitness.  This is especially true for external fertilizers where spawning conditions experienced by eggs will vary dramatically within and among spawning events.   Availability of sperm can vary from scarce to dangerously high concentrations and may depend not only on male abundance and proximity \citep{levitan2002} but also the relative proximity to competing eggs \citep{marshall2005}.  Evidence for fitness effects of both sperm limitation and overabundance include both selective pressure on gamete traits related to fertilization and plasticity in those traits.  For example frequency of alleles associated with egg selectivity may have changed in association with average population density in the red urchin \emph{Strongylocentrotus franciscanus} \citep{levitan2012}.  In addition, gamete traits appear to be directly controlled by both males and females depending upon the density of spawning adults in the tunicate \emph{Styela plicata} \citep{crean2008}.   Finally, competition among eggs for sperm can provide an explanation for diverse modes of fertilization across marine taxa \citep{henshaw2014}. 

In this study, the effect of egg concentration on fertilization presents a potential additional source of selective pressure.  Releasing too many eggs under conditions of low sperm concentrations can further reduce the already low probability of fertilization.   In contrast releasing fewer eggs at high sperm concentrations heightens the risk of polyspermy by increasing the rate of per capita sperm-egg collisions.   The nonlinear sperm-egg interactions may provide another behavioral source of selection in terms of egg release rate if these interactions also exist in nature.   Under this model, females locally exposed to high concentrations of sperm during a spawning event may benefit from releasing many eggs rapidly to minimize polyspermy.  In contrast, females surrounded by low sperm concentrations may benefit from releasing fewer eggs over longer durations or over multiple spawning events to maximize fertilization probability.   Yet how predictions play out in nature remains speculative because the mechanism behind the nonlinear per capita interaction rates are at present unknown as this study presents the first evidence of such dynamics.  
 
\begin{figure*}[t]
\begin{center}
\includegraphics[width= 120mm]{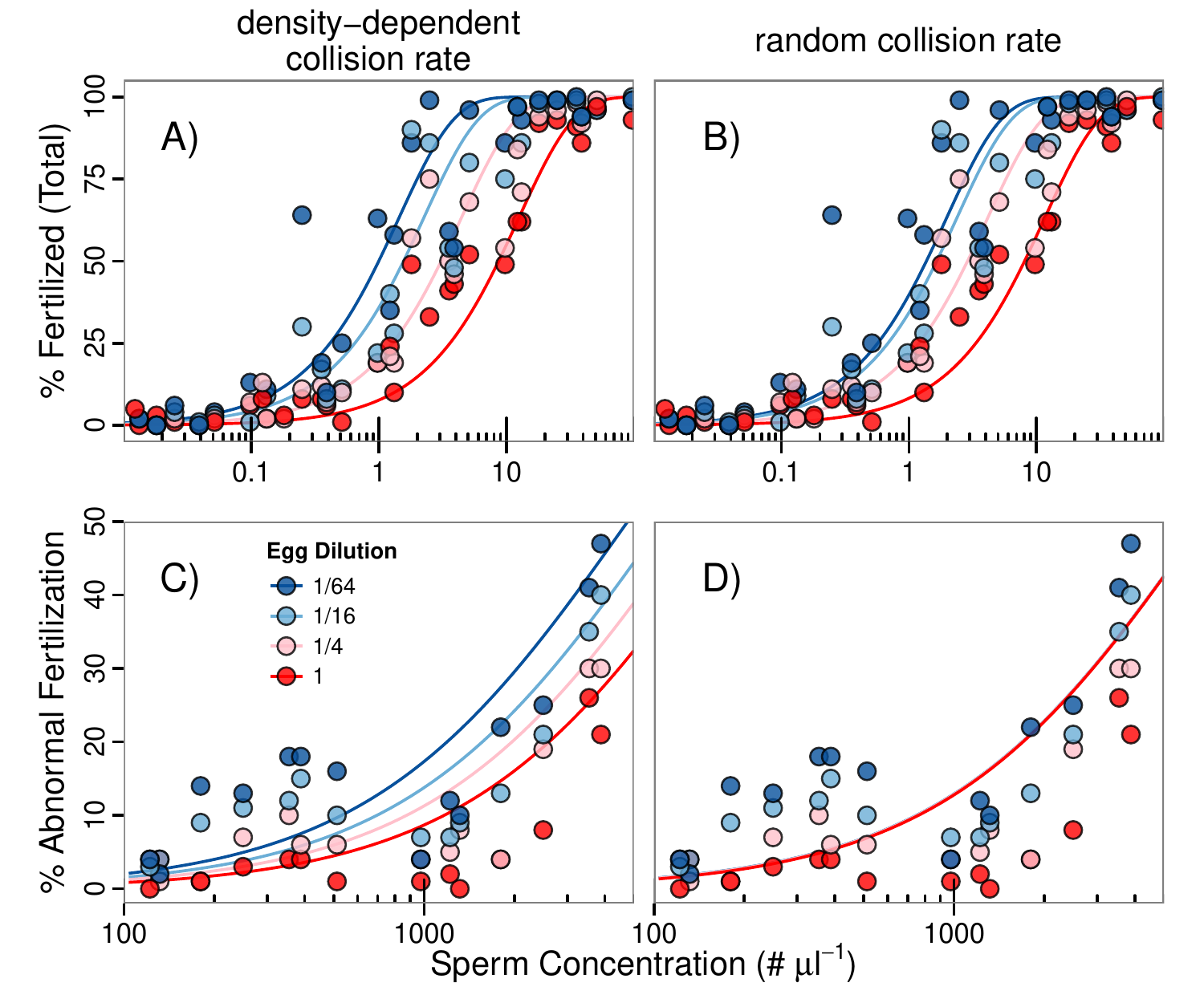}
\end{center}
\caption[Model predictions under different hypotheses]{Model predictions under nonlinear collision rates (A \& C) or random collision rates (i.e. collision rates held constant) (B \& D) for \% total fertilization (A \& B) and \% abnormal fertilization (a proxy for polyspermy) (C \& D). Lines represent the among pair mean expectation from the dynamic model posterior.   Panels A \& C represent the same predictions as Figure \ref{fig:F3} parsed into total fertilization and polyspermy.  Colors indicate different egg concentrations, with blue representing lower concentrations.  }
\label{fig:F4}
\end{figure*}

Accumulation of sperm near eggs is a plausible mechanism for the observed nonlinear per capita sperm-egg interactions.  This can arise directly from attraction.  In their original work \cite{vogel1982} discounted the possibility of attraction of sperm by eggs.   Others, however, have subsequenty demonstrated occurrence of sperm chemotaxis in broadcast spawners \citep{kaupp2012,riffell2004}.     To my knowledge there are not any studies that provide an explicit demonstration of attraction via directional chemotaxis in \emph{S. purpuratus} in the laboratory.  Yet observations of sperm aggregating around individual eggs in \emph{S. purpuratus} date back more than a century \citep{elder,loeb1914} and a primary commercial chemoattractant (speract) derived from \emph{S. purpuratus} eggs \citep{speract} is capable of altering the behavior and swimming speed of \emph{S. purpuratus} sperm \citep{wood2007}.  Moreover, directional chemoattraction is not needed to yield accumulation of sperm around eggs \citep{jaiswal1999}.   In addition to attraction, other mechanisms may explain the observation that per capita collision rates decline with egg concentration.  For example, egg crowding effects that interfere with sperm search patterns may also explain the pattern consistent with these observations.  More thorough evaluation of the dynamic behavior and fate of sperm through time may prove essential for testing hypotheses under both simple and more complex circumstances.  Regardless of the mechanism that leads to this change in per capita collision rates, my results illustrate that the assumptions of random collisions and minimal impact of egg concentrations on fertilization are, in this case, invalid.   

Models include simplifying assumptions for tractable purposes, yet inferences gleamed from simplistic models parameterized with laboratory data remain partly constrained by the unique context of the study.   Historical use of simple fertilization kinetics models includes comparison of parameters  across taxa, across experimental treatments and extrapolation of parameters to the field.   In such cases, experimental analysis should carefully examine sensitivity of results to relaxation of those assumptions.  In controlled laboratory studies, water motion is generally minimal and sperm and eggs are introduced simultaneously.  In nature, spawning events can occur in a dynamic fluid environment where sperm and eggs from many parents meet in turbulent or high velocity circumstances \citep{denny1989,thomas2013}.   Flow and shear stress can inhibit the ability of sperm to directionally seek eggs and alter fertilization rates \citep{riffell2007,zimmer2011} and turbulence may overwhelm such mechanisms \citep{denny1989}.   Thus, if accumulation of sperm around eggs is the process that results in non-random collision rates, then such effects will likely be relaxed if sperm behavior is restricted by physical conditions.  Moreover, sperm often exhibit circular or patterned search behavior \citep{farley2002} that can induce complex vortexes and the hydrodynamic conditions controlling such behavior likely differ in the ocean.  Such differences include the dispersal of chemoattractive gradients, shear stress that impedes swimming speed and direction, or heterogeneity in sperm concentrations where wisps of sperm rather than broad clouds mix with pockets of eggs in suspension.   For example, natural spawning events extend beyond simple reaction-diffusion dynamics as sperm and eggs are released in viscous clouds that create strong spatial heterogeneity in both gamete concentration and fluid dynamics \citep{thomas1994}.  Simple laboratory experiments where gametes are homogenized by agitation can minimize such effects, but characterizing natural spawning events may require more sophisticated experiments as well as models that can capture such complex dynamics.  In addition to the challenges in extrapolating from the laboratory to the field, comparisons across taxa, even among congeners can be dangerous.   For example fertilization dynamics in \emph{Strongylocentrotus franciscanus}, the congener of \emph{S. purpuratus}, showed no sensitivity to egg concentration in the laboratory \citep{levitan1991}.  However, such observations may arise from the small range of egg concentrations used.  Despite these concerns, these results clearly indicate that egg concentrations can, in some taxa, play an important role in altering the nature of fertilization dynamics.   Such nonlinearities that alter per-capita rates of interaction in dynamic systems can drastically change expected outcomes and trajectories.  This is evident for different functional responses in predator-prey systems  \citep{cosner1999} or disease transmission \citep{mccallum2001}.   As such, importance of these observed shifts in rates of interaction to questions of fitness should be tested and explored in future studies.   The observations and model derived here, building upon the advances of \cite{vogel1982}, \cite{styan1998} and \cite{millar2003}, can provide a starting point for expanding theoretical and empirical inquiry into realistic settings regarding if and how such nonlinearities impact fertilization and fitness.  

Existing models of fertilization ignore the presence of nonlinear sperm-egg collision rates that can impact fitness.  These dynamics may empower females with additional behavioral controls that can buffer against both under-fertilization (because of too few sperm) and polyspermy (too many sperm).  Fewer eggs released when sperm are sparse can both increase per capita collision rates and decreases the number of sperm depleted, thereby increasing fertilization probabilities.  More eggs released when sperm are abundant decreases per capita collision rates and buffers against polyspermy.  In addition, proximity to other females during spawning contributes to the pool of competing eggs.   As a result fertilization rates and reproductive efficiency may be controlled not only by the timing of egg release \citep{levitan2005} but the rate of egg release and proximity to competing females as well.    These potential controls arise from the trade-off illustrated here: as sperm concentrations increase, the effect of eggs on one another transitions from competition for sperm to cooperation.  

\section*{Acknowledgements}
Funding for this research was provided by the National Science Foundation in support of the Santa  Santa Barbara Coastal Long Term Ecological Research (LTER) site (NSF OCE 1232779). Additional funding sources of funding included fellowships from the UCSB Graduate Division, the UC Affiliates, and the Department of Ecology, Evolution and Marine Biology at UCSB.  Don Levitan provided critical insight through both discussions and valuable comments on earlier versions of the manuscript.  I also thank Jeff Riffell, Richard Zimmer and Roger Nisbet for taking the time for discussions and Dan Reed, Sally Holbrook, Russell Schmitt and Cherie Briggs for their important criticism on earlier versions of the manuscript.  Finally, I am thankful to Dustin Marshall, Yannis Michalakis, Michael Neubert and one anonymous reviewer for insightful critiques that substantially improved the clarity and quality of this manuscript.

\begingroup
\bibliographystyle{amnat.bst}
\bibliography{UrchinFert.bib}
\clearpage
\endgroup

\clearpage
\pagenumbering{arabic} 
\renewcommand{\thepage}{A\arabic{page}}
\title{Appendices for ``Competition among eggs shifts to cooperation along a sperm supply gradient in an external fertilizer"}
\nolinenumbers
\date{}
\maketitle

\newpage
\setcounter{table}{0}
\nolinenumbers
\renewcommand{\thetable}{A.\arabic{table}}
\setcounter{figure}{0}
\renewcommand{\thefigure}{A.\arabic{figure}}
\setcounter{equation}{0}
\renewcommand{\theequation}{A.\arabic{equation}}
\rhead{Appendix A: alternative model forms}
\section*{Online Appendix A: forms and complete estimates for alternative model forms}

The models of \cite{styan1998} and \cite{millar2003} are given by:
\begin{align}
x(t)&=-\gamma\frac{S_0}{E_T}(1-\exp(-t\beta{E_T}))\\
E_M(t)&=1-e^{-x(t)}-[1-e^{-x(t)}-x(t)e^{-x(t)}][1-e^{-x(t_b)}]
\eqname{Styan (1998)}\\
E_M(t)&=[x(t)-x(t-t_b)]e^{-x(t)}-[e^{-x(t)}-e^{-x(t)}]e^{\beta E_T t_b}]
\eqname{Millar \& Anderson (2003)}
\end{align}
Where parameters and variables match those in Table \ref{table:parametertable} and $t_b$ is the time required between the first sperm entering an egg and the polyspermy block.  Posterior means and credible sets for these models are given in Tables \ref{table:param.values_S} and \ref{table:param.values_MA}.   

When incorporating nonlinear collision rates both model forms show improved WAIC over models with random sperm-egg collision rates (WAIC of 1315.5 versus 1354.7 respectively in the Styan model and WAIC of 1316.2 versus 1355.8 respectively in the MA model).   

Parameter estimates for ${{\beta}}_{\frac{1}{64}}$ vary directly with the input value of $\tau$, but other parameter estimates remain largely invariant.  Thus, comparison with previous estimates for collision rates should consider that 1) attack rates depend directly on egg concentration 
and 2) the value is sensitive to $\tau$.  
\clearpage
\begin{table}[ht]
\caption{Full table of parameters, description, estimated posterior mean, 95\% credible set, and the prior for the dynamic model with unique collision rates (U = uniform prior).}
\centering
 \begin{threeparttable}
\begin{tabular}{ccccc}
\hline
Parameter & Description & Mean & 95\% Credible Set & Prior\\
\hline
${{\beta}}_{\frac{1}{64}}$&  median\mtnote{$\dagger$} collision rate ($\frac{1}{64}$) & 2.59$\times {10}^{-3}$ & (0.99 : 7.62)$\times {10}^{-3}$ & U[1$\times{10}^{-5} : 0.1]$\\
${{\beta}}_{\frac{1}{16}}/{{\beta}}_{\frac{1}{64}}$ & collision rate ratio & 0.74 & 0.56 : 0.98 & U$[0.01 : 2.00]$\\
${{\beta}}_{\frac{1}{4}}/{{\beta}}_{\frac{1}{64}}$ & collision rate ratio & 0.59 & 0.42 : 0.84 & U$[0.01 : 2.00]$\\
${{\beta}}_{1}/{{\beta}}_{\frac{1}{64}}$ & collision rate ratio & 0.45 & 0.30 : 0.64 & U$[0.01 : 2.00]$\\
${\gamma}$ & mean egg selectively & 0.081 & 0.075 : 0.129  & U[1$\times{10}^{-5} : 0.15]$\\
${\delta}$& media\mtnote{$\dagger$} n polypsermy block rate & 1.84 & 1.00 : 3.26 & U[1$\times{10}^{-5}: 20.0]$\\
$\lambda$ & total count ($\text{dispersion}^{-1}$) & 31.1 & 23.8 : 39.4 & Pareto$[1,1.5]$\\
\\
$\sigma_{\beta_\frac{1}{64}}$ & sd in collision rate\mtnote{$\ddagger$} & 1.15 & 0.57 : 2.18 & U[1$\times 10^{-5} : 4]$\\
$\sigma_\gamma$ & sd in egg selectively & 1.21 & 0.02 : 3.84  & U[1$\times{10}^{-5} : 4]$\\
$\sigma_{\delta}$ & sd in polyspermy block\mtnote{$\ddagger$} & 1.05 & 0.58 : 2.09 & U[1$\times{10}^{-5} : 4]$\\
\hline\end{tabular}
\begin{tablenotes}
\footnotesize
            \item[$\dagger$] Among-pair median values are reported because the among-pair hierarchical distribution is assumed to be lognormal and thus the exponentiated posterior mean equals the median.
            \item[$\ddagger$] log scale due to lognormal distribution.
        \end{tablenotes}  
 \end{threeparttable}
\label{table:param.values_Dyn}
\end{table}
\clearpage

\begin{table}[ht]
\caption{Table of parameters, description, estimated posterior mean, 95\% credible set, and the prior for the modified model of \cite{styan1998} with unique collision rates (U = uniform prior). }
\centering
 \begin{threeparttable}
\begin{tabular}{ccccc}
\hline
Parameter & Description & Estimate & 95\% Credible Set & Prior\\
\hline
${{\beta}}_{\frac{1}{64}}$& median\mtnote{$\dagger$} collision rate ($\frac{1}{64}$) & 2.69$\times {10}^{-3}$ & (1.27 : 5.89)$\times {10}^{-3}$ & U[1$\times{10}^{-5} : 0.1]$\\
${{\beta}}_{\frac{1}{16}}/{{\beta}}_{\frac{1}{64}}$ & collision rate ratio & 0.75 & 0.59 : 0.93 & U$[0.01 : 2.00]$\\
${{\beta}}_{\frac{1}{4}}/{{\beta}}_{\frac{1}{64}}$ & collision rate ratio & 0.54 & 0.41 : 0.73 & U$[0.01 : 2.00]$\\
${{\beta}}_{1}/{{\beta}}_{\frac{1}{64}}$ & collision rate ratio & 0.41 & 0.31 : 0.56 & U$[0.01 : 2.00]$\\
$\gamma$ & mean egg selectively & 0.083 & 0.075 : 0.133  & U[1$\times{10}^{-5} : 0.15]$\\
$\delta$ & median\mtnote{$\dagger$} polypsermy block rate & 1.63 & 0.68 : 3.71 & U[1$\times{10}^{-5}: 20.0]$\\
$\lambda$ & total count ($\text{dispersion}^{-1}$) & 31.5 & 23.9 : 40.3 & Pareto$[1,1.5]$\\
\\
$\sigma_{\beta_\frac{1}{64}}$ & sd in collision rate\mtnote{$\ddagger$} & 0.89 & 0.41 : 1.79 & U[1$\times 10^{-5} : 4]$\\
$\sigma_\gamma$ & sd in egg selectively & 8.74 $\times {10}^{-2}$ & (0.39 - 14.89)$\times {10}^{-2}$  & U[1$\times{10}^{-5} : 4]$\\
$\sigma_{\delta}$ & sd in polyspermy block\mtnote{$\ddagger$} & 1.16 & 0.53 - 2.51 & U[1$\times{10}^{-5} : 4]$\\
\hline\end{tabular}
\begin{tablenotes}
\footnotesize
            \item[$\dagger$] Among-pair median values are reported because the among-pair hierarchical distribution is assumed to be lognormal and thus the exponentiated posterior mean equals the median.
            \item[$\ddagger$] log scale due to lognormal distribution.
        \end{tablenotes}  
 \end{threeparttable}
\label{table:param.values_S}
\end{table}

\clearpage
\begin{table}[ht]
\caption{Table of parameters, description, estimated posterior mean, 95\% credible set, and the prior for the modified model of \cite{millar2003} with unique collision rates (U = uniform prior).}
\centering
 \begin{threeparttable}
\begin{tabular}{ccccc}
\hline
Parameter & Description & Estimate & 95\% Credible Set & Prior\\
\hline
${{\beta}}_{\frac{1}{64}}$ & median\mtnote{$\dagger$} collision rate ($\frac{1}{64}$) & 2.59$\times {10}^{-3}$  & (1.25 : 5.24)$\times {10}^{-3}$ & U[1$\times{10}^{-5}: 0.1]$\\
${{\beta}}_{\frac{1}{16}}/{{\beta}}_{\frac{1}{64}}$ & collision rate ratio & 0.74 & 0.58 : 0.92 & U$[0.01: 2.00]$\\
${{\beta}}_{\frac{1}{4}}/{{\beta}}_{\frac{1}{64}}$ & collision rate ratio & 0.54 & 0.41 : 0.70 & U$[0.01: 2.00]$\\
${{\beta}}_{1}/{{\beta}}_{\frac{1}{64}}$ & collision rate ratio & 0.41 & 0.30 : 0.57 & U$[0.01:2.00]$\\
$\gamma$ & mean egg selectively & 0.087 & 0.006 : 0.147  & U[1$\times{10}^{-5}: 0.15]$\\
$\delta$ & median\mtnote{$\dagger$} polypsermy block rate & 1.02 & 0.41 : 2.41 & U[1$\times{10}^{-5}: 20.0]$\\
$\lambda$ & total count ($\text{dispersion}^{-1}$) & 30.5 & 23.8 : 38.6 & Pareto$[1,1.5]$\\
\\
$\sigma_{\beta_\frac{1}{64}}$ & sd in collision rate\mtnote{$\ddagger$} & 0.85 & 0.44 : 1.66 & U[1$\times 10^{-5}: 4]$\\
$\sigma_\gamma$ & sd in egg selectively & 1.65 & 0.02 : 3.87   & U[1$\times{10}^{-5}: 4]$\\
$\sigma_{\delta}$ & sd in polyspermy block\mtnote{$\ddagger$} & 1.04 & 0.52 : 1.95  & U[1$\times{10}^{-5}: 4]$\\
\hline\end{tabular}
\begin{tablenotes}
\footnotesize
            \item[$\dagger$] Among-pair median values are reported because the among-pair hierarchical distribution is assumed to be lognormal and thus the exponentiated posterior mean equals the median.
            \item[$\ddagger$] log scale due to lognormal distribution.
        \end{tablenotes}  
 \end{threeparttable}
 \label{table:param.values_MA}
\end{table}
\clearpage

\rhead{Appendix B: sperm ``half-life" and decay rates}
\section*{Online Appendix B: sperm ``half-life" and decay rates}
\setcounter{table}{0}
\renewcommand{\thetable}{B.\arabic{table}}
\setcounter{figure}{0}
\renewcommand{\thefigure}{B.\arabic{figure}}
\nolinenumbers

The sperm ``half-life'' is described by \cite{vogel1982} as the time after which total fertilization capacity of sperm in solution is reduced to 50\% of the capacity at release.   However, the fertilization capacity of sperm does not necessarily linearly track the degradation of sperm (i.e. a change in the population of viable sperm does not necessarily lead to a proportional change in fertilization rates).   However, it is possible that the half-life of sperm can be independent of sperm concentration but fertilization half-life is not.  The latter was observed by  \cite{levitan1993} and \cite{levitan1991}.  In this case, the latter will represent a biased approach.

\begin{figure}
\begin{center}
\includegraphics{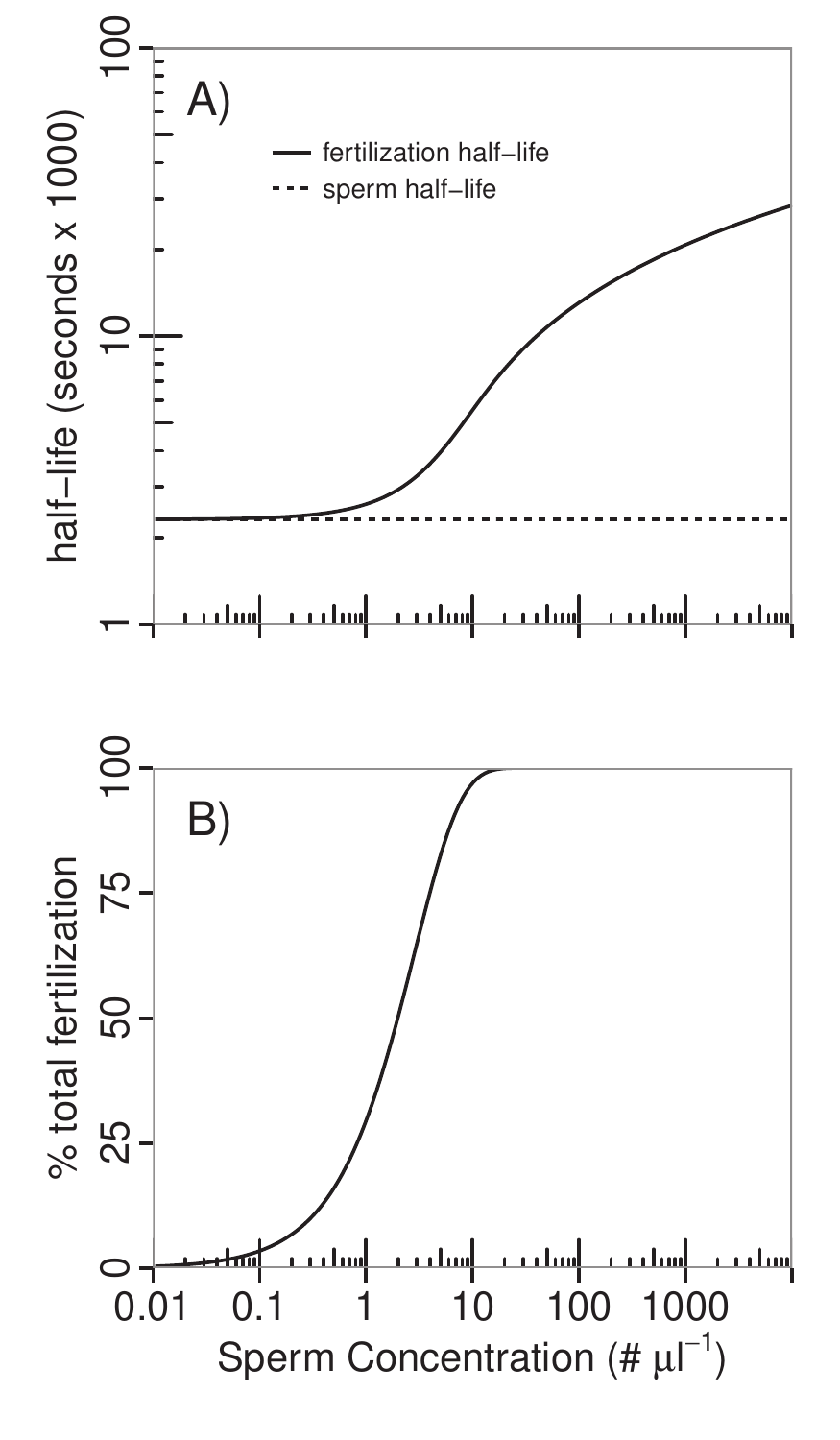}
\end{center}
\caption[Bias in estimates of sperm-half life]{A) Half-life values produced under a constant sperm decay rate.   The sperm half-life (dotted) is constant while the fertilization half-life (solid line) of \cite{vogel1982} is biased towards higher values, increasing as sperm concentrations and fertilization rates (B) increase.    }
\label{fig:FA1}
\end{figure}

Let the definition of \cite{vogel1982} be called ``fertilization half-life" and the half-life of sperm be called ``sperm half-life".    The exact bias of the \cite{vogel1982} method will differ depending upon the functional form of sperm degradation.   Consider a constant decay rate as in eq. \eqref{eq:1}.  In this case percent fertilization at a given time is $1- E_U(t)/E_U(t=0)$, where $E_U(t)$ is provided by eq. \eqref{eq:4b}.  In this case, the sperm half-life is given by $\log(0.5)/r$.   If one uses parameter estimates from the model estimated below (with constant egg concentrations) the sperm-half life is constant but the fertilization half-life is biased and increases with sperm concentration (Fig. \ref{fig:FA1}).  

Thus using the definition of \cite{vogel1982} will naturally produce biased estimates if the decay function is independent of sperm concentration.   To produce unbiased estimates, one must 1) estimate the empirical relationship between sperm concentrations and fertilization rates for virgin sperm at age $\approx 0$;  2) conduct a factorial age-concentration experiment; 3) use only sperm concentrations that produce less than 100\% fertilization at age $\approx 0$; 4) for each result calculate the number of sperm required to achieve such fertilization; 5) estimate the loss of sperm at each age within each concentration; and 6) estimate the age at which 50\% of initial sperm concentrations are lost.    A better solution is to directly estimate the decay rate and use a model that explicitly includes decay rate, such as eqns. \eqref{eq:1}-\eqref{eq:5}.

\newpage

\subsection*{Experimental estimation of ``half-life" $\tau$ and decay rate $r$}
In order to estimate the sperm decay rate ($r$) and half-life, I conducted an experiment identical to that described above, but used only one egg concentration (assuming egg concentration does not affect the intrinsic decay rate), adding sperm to vials containing eggs at $t\in 0, 600, 1200, 2700, 3600 \text{ and } 7200$ seconds after initial sperm release and dilution for each of four sperm dilutions ($10^{-4}, 10^{-5}, 10^{-6},\text{ }\&\text{ } 10^{-7}  \text{ }\mu l^{-1}$).   I excluded sperm concentration treatments that started with 100\% fertilization for age 0 sperm.  For each I used the fertilization fraction and the empirical fertilization-sperm concentration relationship (see figure \ref{fig:F1}) to calculate the expected initial sperm concentration.  From these curves I then calculated both the expected decay rate and the ``half-life".  In addition to a constant decay rate, I also estimated other decay functions including: Gompertz, logistic, log-logistic and decay rates that vary by sperm concentration.  Each function was estimated using nonlinear least squares and compared using AICc (corrected Akaike Information Criterion).   

\begin{table}[ht]
\caption[Comparison of different sperm decay functions]{Table of model comparison of nonlinear decay functions.  Lower AICc values are favored and the model within 2 AICc units of the minimum is preferred.}
\begin{center}
\begin{tabular}{ccccc}
\hline
Model & AICc & df & $\Delta$AICc & AICc weight\\
\hline
\textbf{constant decay} & \textbf{29.9} & 2 & \textbf{0.1} & \textbf{0.36}\\
log-logistic & 29.8 & 3 & 0.0 & 0.38\\
Gompertz & 31.7 & 3 & 1.9 & 0.15\\
decay varies by sperm concertration & 32.2 & 4 & 2.4 & 0.11\\
logistic & 41.2 & 2 & 11.5 & 0.01\\

\hline\end{tabular}
\end{center}
\label{table:AICc.decay.values}
\end{table}

There is no support from these data for a decay rate that varies by sperm concentration.  However, these data are designed to estimate an exponential decay rate rather than to test for such differences and thus may lack sufficient power.   From the constant decay function the decay rate is estimated to be $r= 2.7 \times 10^{-4}$ per second ($\pm 0.34\times 10^{-4}$), from which the ``half-life" is estimated to be 2310.5 seconds. 

\newpage
\rhead{Appendix C: hierarchical parameters}
\section*{Online Appendix C: hierarchical parameters and correction for moments of truncated distributions}
\renewcommand{\theequation}{C.\arabic{equation}}
\setcounter{equation}{0}

For ${\beta}$ and ${\delta}$ the data were sufficient to inform the posterior and thus I assumed the pair-level parameters conformed to a lognormal distribution.   In contrast, for ${\gamma}$ the data were not very informative, and thus I assumed the the pair level estimates conformed to a normal distribution for simplicity.   Because I put constraints on the pair level estimates (to improve convergence), all of these distributions are therefore truncated.  I report the uncorrected posterior for comparison with the prior which are close to the empirical means of the posterior pair-level parameters.  If one is concerned with among pair inference, the standard moments of a full distribution are biased when the distribution is truncated and one can easily correct for truncation in the means.  Thus, one should use the estimates provided in Appendix B with the formulas below to generate the unbiased mean and SE for hierarchical moments. 

For a given normal distribution with mean $\mu$ and the standard deviation $\sigma$, the unbiased mean and standard deviation for the truncated distribution are given by:

\begin{align}
\text{mean}_{truncated}&=\mu-\sigma\frac{\phi(x_U)-\phi(x_L)}{\Phi(x_U)-\Phi(x_L)}\\
\text{sd}_{truncated}&=\sigma\sqrt{1-\frac{x_U\phi(x_U)-x_L\phi(x_L)}{\Phi(x_U)-\Phi(x_L)}-\left(\frac{\phi(x_U)-\phi(x_L)}{\Phi(x_U)-\Phi(x_L)}\right)^2}\\
x_U&=\frac{\text{upper bound}-\mu}{\sigma}\\
x_L&=\frac{\text{lower bound}-\mu}{\sigma}
\end{align}
where $\phi(x)$ represents the standard normal probability density and $\Phi(x)$ represents the standard normal cumulative density.

The among pair mean and standard deviation for collision rates (${{\beta}}_{\frac{1}{64}}$) and polyspermy block rate ($\gamma$) were both modeled using a truncated lognormal distribution.   For a given lognormal distribution with median $\mu$ and log-scale standard deviation $\sigma$, the unbiased mean and standard deviation for the truncated distribution are given by:
	
\begin{align}
\text{mean}_{truncated} &=\mu e^{\frac{\sigma^2}{2}}\left[\frac{\Phi(\sigma-x_L)-\Phi(\sigma-x_U)}{\Phi(x_U)-\Phi(x_L)}\right]\\
\text{sd}_{truncated}&=\sqrt{e^{2\ln(\mu)+2\sigma^2}\left[\frac{\Phi(2\sigma-x_L)-\Phi(2\sigma-x_U)}{\Phi(x_U)-\Phi(x_L)}\right]}\\
x_U&=\frac{\ln(\text{upper bound})-\ln(\mu)}{\sigma}\\
x_L&=\frac{\ln(\text{lower bound})-\ln(\mu)}{\sigma}
\end{align}

 \rhead{Appendix D: Example extension to reaction-diffusion problems}
 \section*{Online Appendix D: Example extension to reaction-diffusion problems}
 \setcounter{table}{0}
\renewcommand{\thetable}{D.\arabic{table}}
\setcounter{figure}{0}
\renewcommand{\thefigure}{D.\arabic{figure}}
\setcounter{equation}{0}
\renewcommand{\theequation}{D.\arabic{equation}}
To illustrate the application of the updated dynamic model towards spatial problems, consider the following simple, heuristic example  illustrated in Fig. \ref{fig:pde_sim}.  The reaction equations \eqref{eq:1} - \eqref{eq:5} can be converted into reaction diffusion equations.  

\begin{subequations}
\begin{align}
\frac{\partial S}{\partial t} &=-{\beta} S {E}_{T}-rS+ D_S \nabla^2 S\\
\frac { { { \partial E }_{U} } } {\partial t} &=-\gamma {\beta}  S{E}_{U}+ D_E \nabla^2 {E}_{U}\\
\frac{{{\partial E}_{V}}}{\partial t} &=\gamma {\beta}  S{E}_{U}-\delta{E}_{V}-\gamma {\beta}  S{E}_{V}+D_E \nabla^2 {E}_{V}\\
\frac{{{\partial E}_{M}}}{\partial t} &=\delta{E}_{V}+D_E \nabla^2 {E}_{M}\\
\frac{{{\partial E}_{P}}}{\partial t} &=\gamma {\beta}  S{E}_{V}+D_E \nabla^2 {E}_{P}
\end{align}
\end{subequations}
Where $D_S$ and $D_E$ are the diffusion constants for sperm and eggs which respectively scale the diffusive flux.  

These equations can now be used to evaluated whether eggs closer to a source of sperm can impede fertilization of eggs further away.   Consider that either one or two clouds of eggs are released into a closed environment where  $D_E = 0$ (i.e. eggs do not diffuse)  with no directional advection.   Sperm is introduced at a single point and in the two egg cloud case, one lies closer to this source.   For simplicity the system is represented as a two dimensional 151 x 51 matrix (with arbitrary units).   The egg clouds have a radius of 25 arbitrary units at a concentration of 1 $\mu l^{-1}$ and sperm are introduced in a single unit a concentration of $1\times10^5 \mu l^{-1}$ and diffuse in two dimensions with a diffusion constant of $D_S$ = 0.5 units $s^{-1}$.   

The solutions for the two scenarios are presented in the animations in Fig. \ref{fig:pde_sim}.   In the two egg cloud scenario (left column, black lines) eggs closer to the source deplete available sperm prior to reaching the second egg cloud sufficient to depress fertilization compared to the single egg cloud scenario (middle column, grey lines).   Note also the small area of polyspermy near the origin of the introduction of concentrated sperm.  Simulation results presented are for heuristic purposes only and outcomes will depend heavily on model parameter values.  

The system was solved by discretizing into spatial units and converting the PDEs back into a matrix of ODEs by numerical differencing.  Because this example is presented for heuristic purposes only, there are no complex fluid dynamics and the only boundary conditions introduced are that flux at the matrix margins is zero.   

\newpage
 \begin{figure}[ht]
  \caption{Animated simulation of the reaction-diffusion equations in two dimensions through time where two (A-D) or one (E-H) egg clouds are exposed to sperm along with time series of solutions (I - L) at the arrows for each panel.  Sperm is introduced at a single point to the right of the egg clouds at $1\times10^5 \mu l^{-1}$ with diffusion constants of $D_S$ = 0.5 units $s^{-1}$.  In this case, eggs are not allowed to diffuse and are static at a concentration of 1 $\mu l^{-1}$.  Model parameters used are the median values represented in Table \ref{table:param.values}.  A screenshot of the final solution is shown below. } 
 \label{fig:pde_sim}
  \end{figure}
 \includegraphics[width= 6.5in]{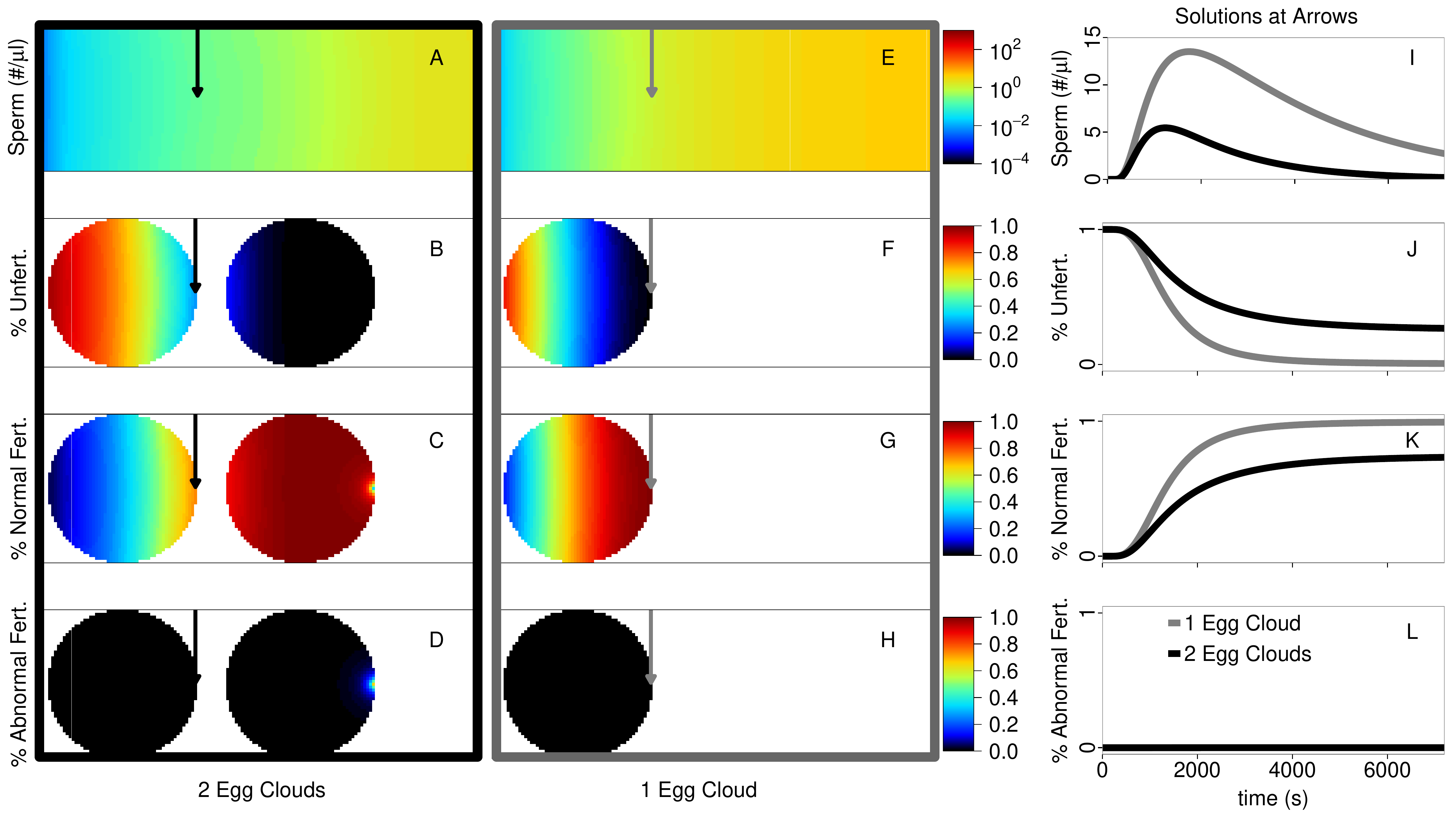}
\end{document}